\documentclass[%
reprint,
superscriptaddress,
amsmath,amssymb,
aps,
pra,
nofootinbib 
]{revtex4-2}
\usepackage{tabularray}
\usepackage{graphicx}
\usepackage{dcolumn}
\usepackage{bm}
\usepackage[colorlinks=true,linkcolor=red,citecolor=blue,urlcolor=blue,anchorcolor=blue,hypertexnames=false]{hyperref}

\usepackage{mathptmx}

\newcommand{\taumathptmx}{%
    \mathchoice
        {\text{\usefont{OML}{ptmcm}{m}{it}\symbol{"1C}}} 
        {\text{\usefont{OML}{ptmcm}{m}{it}\symbol{"1C}}} 
        {\text{\usefont{OML}{ptmcm}{m}{it}\symbol{"1C}}} 
        {\text{\usefont{OML}{ptmcm}{m}{it}\symbol{"1C}}} 
}

\usepackage{xcolor}         
\usepackage[normalem]{ulem} 
\usepackage{soul}

\makeatletter
\DeclareSymbolFont{letters}{OML}{cmm}{m}{it}
\SetSymbolFont{letters}{bold}{OML}{cmm}{b}{it}
\makeatother
\renewcommand{\tau}{\taumathptmx}
\usepackage{soul, color, xcolor} 
\usepackage{ulem}
\usepackage{xspace}

\newcommand{\addressa}{\affiliation{Laboratory of Quantum Information, University of Science and Technology of China, Hefei 230026, China}}
\newcommand{\addressb}{\affiliation{CAS Center For Excellence in Quantum Information and Quantum Physics, University of Science and Technology of China, Hefei, Anhui, 230026, China}}
\newcommand{\addressc}{\affiliation{Origin Quantum Computing Company Limited, Hefei, Anhui 230026, China}}
\newcommand{\addressd}{\affiliation{Suzhou Institute for Advanced Research, University of Science and Technology of China, Suzhou, Jiangsu 215123, China}}

\begin{document}
    \preprint{APS/123-QED}
    \title{Overcoming the Speed–Fidelity Trade-off in Fast CZ Gates via Cyclic Control}
    
    \author{Ze-An Zhao}\addressa\addressb
    \author{Hai-Feng Zhang}
    \author{Tian-Le Wang}
    \author{Xiao-Yan Yang}
    \addressa\addressb
    \author{Peng Wang}
    \addressa\addressb\addressd
    \author{Ren-Ze Zhao}
    \addressa\addressb
    \author{Sheng Zhang}
    \addressa\addressb\addressd
    \author{Zhi-Fei Li}
    \author{Yuan Wu}
    \author{Zi-Hao Fu}
    \author{Sheng-Ri Liu}
    \addressa\addressb
	\author{Peng Duan}
    \email{pengduan@ustc.edu.cn}
    \thanks{Corresponding author}
    \addressa\addressb

    \author{Guo-Ping Guo}
    \email{gpguo@ustc.edu.cn}
    \thanks{Corresponding author}
    \addressa\addressb\addressc

    \begin{abstract}

    High-fidelity quantum gates are essential for scalable quantum computation. However, at short durations, short-timescale waveform distortions break the time-reflection symmetry of control pulses, preventing the precise closure of cyclic evolution. This mechanism renders conventional symmetric protocols intrinsically over-constrained. Conventional strategies typically rely on smoothing the pulse envelopes or embedding the interaction pulse within a longer qubit pulse to bypass short-timescale distortions, which inevitably leads to a persistent speed–fidelity trade-off.
    To overcome this limitation, we introduce a cyclic control strategy based on parameter-space expansion, which restores controllability by incorporating an additional degree of freedom. We experimentally demonstrate this approach in a superconducting controlled-Z gate, achieving robust suppression of coherent errors without increasing gate duration, reducing the average coherent error from 0.27\% to 0.12\% across multiple two-qubit gates, as validated by cross-entropy benchmarking. Our results establish a general route to fast, high-fidelity cyclic quantum gates beyond the conventional speed–fidelity trade-off.

    \end{abstract}

    \maketitle

    Superconducting quantum computing is gradually approaching critical milestones, including experimental demonstrations of quantum advantage~\cite{arute2019quantum,wu2021strong,zhong2021phase}, reducing logical error rate by increasing code distance~\cite{krinner2022realizing,google2023suppressing,google2025quantum,eickbusch2025demonstration,lacroix2025scaling} as well as using higher-rate qLPDC codes\cite{bravyi2024high,wang2025demonstrationlowoverheadquantumerror}, and evidence of quantum computing's utility in the pre-fault-tolerant regime~\cite{kim_evidence_2023}. Recent advances on Noisy Intermediate-Scale Quantum (NISQ) processors~\cite{cao2023generation,gupta2024encoding,bravyi2024high,terhal2024mobile,guo2024experimental} indicate tangible progress toward practical applications. Achieving these goals requires fast and high-fidelity execution of deep quantum circuits. However, two-qubit gates remain the primary performance bottleneck, typically exhibiting error rates an order of magnitude higher than single-qubit operations~\cite{wu2021strong,google2023suppressing}. In particular, fast entangling gates are often limited by a practical speed–fidelity trade-off: reducing gate duration generally leads to increased coherent errors, hindering efficient circuit execution.

    A broad class of quantum gates relies on cyclic evolution to accumulate geometric phase, including M\o lmer-S\o rensen gates in trapped ions~\cite{ballance2016high,sutherland2023individual} and geometric phase gates in NV centers~\cite{sekiguchi2017optical,nagata2018universal}. In superconducting circuits, controlled-phase (CPhase) gates share this foundation, where the interaction between the computational state \(|11\rangle\) and the noncomputational state \(|20\rangle\) induces a conditional phase through cyclic evolution~\cite{strauch2003quantum,dicarlo2009demonstration}. Among these, baseband flux-controlled CZ gates achieve the fastest operation, approaching the speed limit \(t_{\mathrm{lim}} = \pi / g\)~\cite{barends2019diabatic}, and significantly outperform microwave-driven cross-resonance~\cite{sheldon2016procedure} and parametric gates~\cite{hong2020demonstration}. Recent implementations using tunable coupler architectures further enable strong, controllable interactions with suppressed residual coupling~\cite{li2020tunable,xu2020high,sung2021realization}.

    In realistic devices, however, hardware imperfections disrupt the cyclic evolution underlying these gates. While long-timescale distortions with response timescales exceeding $\sim$10\,ns can be mitigated through calibration and filtering techniques~\cite{foxen2018high,yan2019strongly,rol2020time,rol2019fast,negirneac2021high}, short-term distortions (STDs), with sub-10\,ns timescales, remain difficult to characterize and suppress~\cite{barends2014superconducting,kelly2015state}. 
    Here, we reveal that STDs and pulse timing misalignments fundamentally break the time-reflection symmetry [TFS: $H(t)=H(\tau-t)$] of the control waveforms. 
    This symmetry breaking increases the number of physical constraints required to precisely close the cyclic evolution, rendering conventional symmetric protocols intrinsically over-constrained.
    To mitigate these errors, conventional strategies typically rely on smoothing the pulse envelopes or embedding the coupler interaction within a longer qubit pulse to bypass transient, short-timescale distortions~\cite{sung2021realization,marxer2025above}. However, such approaches merely compensate for the constraint deficit by sacrificing gate speed, inevitably leading to an intrinsic speed--fidelity trade-off.
    Furthermore, while numerical optimization techniques~\cite{sivak2022model,ding2023high,li2024realization} offer a route to high fidelities, they operate as physical black boxes and often lack hardware transferability.

    In this Letter, we propose a universal control strategy based on parameter-space expansion (PSE) to resolve this fundamental imbalance. 
    The PSE protocol introduces an auxiliary degree of freedom to the control waveform. 
    This mathematically satisfies the additional constraints imposed by TFS breaking, thereby suppressing both leakage and phase errors without prolonging the gate evolution. 
    We experimentally demonstrate this principle using CZ gates. 
    Across multiple two-qubit gates, our PSE-CZ scheme consistently reduces distortion-induced coherent errors while strictly maintaining the original gate duration. 
    Ultimately, the identified symmetry-breaking mechanism and our PSE solution apply broadly to general cyclic quantum control, including geometric and holonomic gates, providing a definitive route beyond current hardware limitations.

\begin{figure}[t]
    \centering
    \includegraphics[width=1\linewidth]{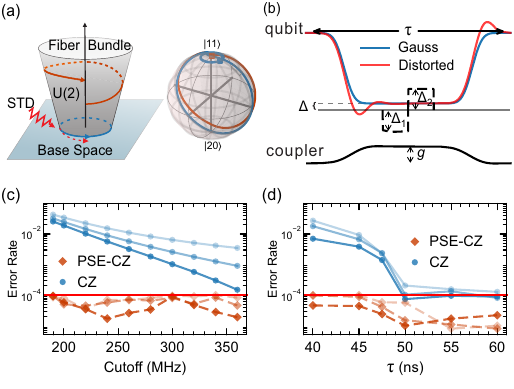}
    \caption{ \textbf{Geometric mechanism and simulation of the PSE-CZ gate.} (a) Left: Cyclic evolution in the base space induces a non-Abelian geometric phase in the $\mathrm{U}(2)$ bundle, which is susceptible to hardware imperfections such as STD. Right: Bloch sphere trajectories in the \(|11\rangle-|20\rangle\)  subspace for a non-cyclic gate resulting from STD (blue) and an ideal cyclic gate (orange); squares denote evolution endpoints. (b) CZ gate pulse sequence (duration \(\tau\)) utilizing flattop Gaussian waveforms with distortions simulated via a Butterworth low-pass filter. The PSE-CZ scheme employs independent amplitudes $\Delta_1$ and $\Delta_2$ for the two segments split at the midpoint, with experimental values typically in the range of 1--5~MHz (all frequencies denote $f = \omega/2\pi$). (c) Simulated distortion-induced error rates for CZ and PSE-CZ gates versus filter cutoff frequency. Darker colors denote smaller distortion amplitudes. (d) Simulated gate error rates as a function of duration $\tau$. To maintain a constant effective coupler duration, the coupler buffer is adjusted while keeping the qubit buffer fixed.}
        \label{fig:1}
    \end{figure}

    In standard implementations, the multi-level dynamics of the two coupled transmon qubits reduce to an effective two-level subspace spanned by $|11\rangle$ and $|20\rangle$ (see Supplemental Material~G). 
    The effective Hamiltonian is $H(t)=\frac{1}{2}\hbar g(t)\sigma_x - \frac{1}{2}\hbar\Delta(t)\sigma_z$, which is traditionally controlled by only two degrees of freedom: the coupling strength $g$ and the frequency detuning $\Delta$ at the pulse plateau. 
    Ideally, the CZ gate is realized by a cyclic evolution in this base space. 
    Geometrically, this closed trajectory lifts to a holonomy in the principal $\mathrm{U}(2)$ bundle[Fig.~\ref{fig:1}(a), left], accumulating a non-Abelian geometric phase. 
    Dynamically, this manifests as a fully closed loop on the $|11\rangle$--$|20\rangle$ Bloch sphere (orange), whereas STDs distort this evolution into an unclosed, non-cyclic trajectory (blue)[Fig.~\ref{fig:1}(a), right]. 
    To execute a high-fidelity CZ gate, the evolution must satisfy two fundamental requirements: the leakage condition (\textbf{LC}, $L=0$) to ensure strict cyclicity, and the conditional phase condition (\textbf{PC}, $\mathrm{CPhase} = \pi$).

    Formally, we parameterize the subspace propagator $U(\tau,0) \in \mathrm{SU}(2)$ using Cayley-Klein coefficients $a$ and $b$:
    \[
    U(\tau,0) = \begin{pmatrix} a & b \\ -b^* & a^* \end{pmatrix}, \quad \text{with} \quad |a|^2+|b|^2=1.
    \]
    A cyclic evolution requires a diagonal propagator, forcing the complex off-diagonal element to vanish ($b=0$). 
    This single \textbf{LC} requirement imposes two independent real constraints: $\mathrm{Re}[b]=0$ and $\mathrm{Im}[b]=0$. 
    However, under TFS, the propagator is naturally symmetric ($U=U^{\mathrm{T}}$), which automatically guarantees $\mathrm{Re}[b]=0$. 
    Thus, the \textbf{LC} reduces to a single constraint ($\mathrm{Im}[b]=0$). 
    Combined with the \textbf{PC}, the system exhibits exactly two physical constraints, perfectly commensurate with the two available control parameters ($g$ and $\Delta$).

    Crucially, hardware imperfections such as waveform distortions fundamentally break TFS. 
    This symmetry breaking revives the hidden constraint $\mathrm{Re}[b]=0$, elevating the total requirement to three independent constraints. 
    Consequently, the optimization of a CZ gate with only two control parameters becomes intrinsically over-constrained, rendering the simultaneous satisfaction of \textbf{LC} and \textbf{PC} theoretically impossible with a traditional symmetric pulse. 
    Beyond this two-level subspace, this constraint-counting principle generalizes to arbitrary $D$-level systems: TFS breaking universally increases the number of cyclic constraints from $D(D-1)/2$ to $D(D-1)$, as detailed in Supplemental Material~D.

    To resolve this intrinsically over-constrained problem, we introduce the PSE-CZ protocol. 
    By splitting the detuning profile $\Delta(t)$ into two independently tunable segments ($\Delta_1$ and $\Delta_2$) at the pulse midpoint[Fig.~\ref{fig:1}(b)], we introduce a crucial third degree of freedom. 
    Together with the coupling strength $g$, the parameter set $(\Delta_1, \Delta_2, g)$ exactly matches the three constraints ($\text{Re}[b] = 0$, $\text{Im}[b] = 0$, and $\text{CPhase} = \pi$), restoring the system to an exactly-constrained regime and guaranteeing the existence of cyclic solution.
    To validate this theoretical framework, we numerically evaluate the gate's resilience against STDs. 
    The distortions are modeled by applying a Butterworth low-pass filter to the ideal flat-top Gaussian waveform[Fig.~\ref{fig:1}(b)], with full filter specifications and implementation details provided in Supplemental Material~E.  
    Figure~\ref{fig:1}(c) illustrates the coherent error as a function of the filter cutoff frequency and distortion amplitude. 
    For the traditional CZ gate, the error is highly vulnerable to waveform imperfections, decreasing only when the cutoff frequency increases or the distortion amplitude diminishes. 
    In stark contrast, the PSE-CZ protocol consistently maintains a suppressed coherent error across all simulated cutoff frequencies and distortion amplitudes, demonstrating its inherent robustness against varying STDs.
    Furthermore, the temporal simulations vividly reproduce the speed--fidelity trade-off discussed earlier. 
    As shown in Fig.~\ref{fig:1}(d), the traditional CZ gate exhibits two distinct regimes. 
    For extended durations ($\tau \gtrsim 50$\,ns), a low-error plateau is achieved as qubit distortions are masked by nesting the interaction within a larger qubit pulse window~\cite{marxer2025above}.
    However, as the gate speed increases ($\tau < 50$\,ns), the error surges dramatically because the fixed-timescale distortions increasingly dominate the shortened evolution. 
    The PSE-CZ protocol fundamentally breaks this limitation. 
    It maintains a uniformly low error rate even deep within the fast-gate regime, proving that simultaneous suppression of leakage and phase errors can be accomplished without compromising gate speed.

    \begin{figure}[t]
        \centering
        \includegraphics[width=1\linewidth]{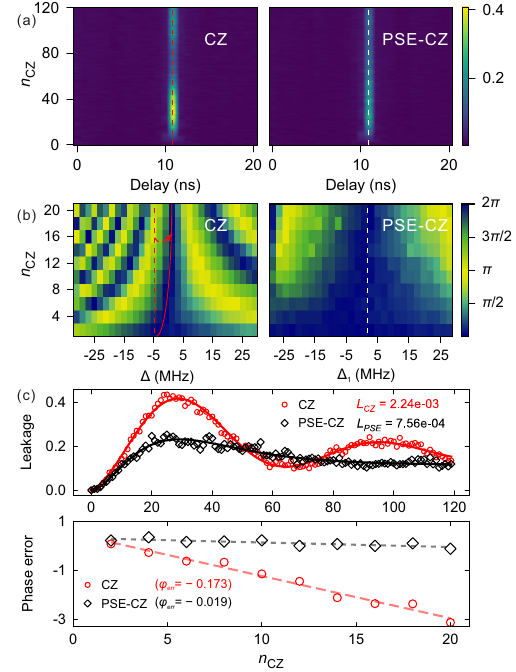}
        \caption{
        \textbf{Experimental diagnosis of non-cyclic evolution and coherent errors.} 
        (a) Coherent amplification of leakage into the \(|20\rangle\) state by repeating the gate with a scanned inter-gate delay, as a function of the number of CZ repetitions $n_{\rm CZ}$.
        (b) Acquired CPhase for (left) traditional CZ and (right) PSE-CZ gates, plotted along the \textbf{LC} solution contour versus $n_{\rm CZ}$. The dashed and solid lines represent the parameters satisfying the \textbf{PC} for $n_{\rm CZ}=1$ and arbitrary \(n_{\rm CZ}\), respectively. Optimal PSE-CZ detunings ($\Delta_1 = 2.6, \Delta_2 = -5.1$\,MHz) are determined via XEB fidelity optimization.
        (c) Single-gate leakage to \(|20\rangle\) (top) and CPhase error (bottom) extracted from (a) and (b). Data points are experimental measurements; solid curves and dashed lines are fits to the data.
        }
        \label{fig:2}
    \end{figure}

    The experiment is performed on a 72-qubit superconducting processor comprising tunable transmon qubits and tunable couplers; relevant device parameters including coherence times and coupling strengths are provided in Supplemental Material~B.
    Building on our predictions, we experimentally diagnose the coherent errors stemming from non-cyclic evolution. We first probe leakage into the $|20\rangle$ state [Fig.~\ref{fig:2}(a)], the most direct signature of non-cyclicity, using an error amplification protocol (see Supplemental Material~F). The single-gate leakage is extracted from the fitted oscillation frequency, which is robust against noise and slow drift owing to the inclusion of $T_1$ and $T_2$ decay in the fitting model. We obtain a leakage of $0.224\%$ for the traditional CZ, which is reduced nearly threefold to $0.0756\%$ for the PSE-CZ gate [Fig.~\ref{fig:2}(c, top)].
    This result provides direct evidence for non-cyclic evolution in the standard protocol and confirms the superior cyclicity of the PSE-CZ gate.
    This non-cyclicity, in turn, manifests as an $n_{\rm CZ}$-dependent CPhase for the traditional gate [Fig.~\ref{fig:2}(b)], leading to a phase error that accumulates nearly linearly with $n_{\rm CZ}$, as shown in Fig.~\ref{fig:2}(c, bottom).
    In stark contrast, the PSE-CZ protocol suppresses this error by an order of magnitude, demonstrating a stable, $n_{\rm CZ}$-independent performance [Fig.~\ref{fig:2}(c, bottom)].
    This experimentally observed non-cyclicity is consistent with the effects of hardware imperfections, such as the imperfectly calibrated FIR filters detailed in Supplemental Material C, which are theoretically shown to break the required symmetries for ideal cyclic evolution. Having established the practical failure of the traditional gate and the effectiveness of our solution, we now elucidate the physical mechanism that enables this robust correction.

    \begin{figure}[t]
        \centering
        \includegraphics[width=1\linewidth]{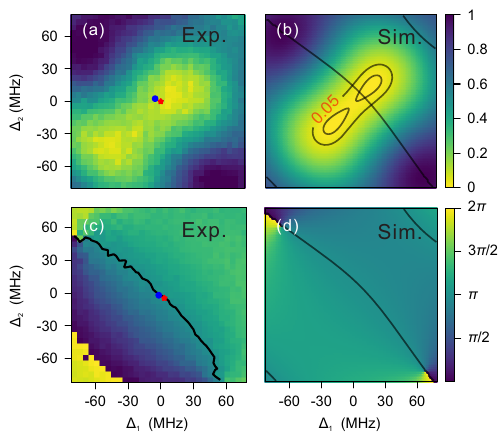}
        \caption{\textbf{Leakage and CPhase landscapes in the expanded parameter space.} (a),(b) Leakage into the \(|20\rangle\) state and (c),(d) CPhase plotted as a function of \(\Delta_{1}\) and \(\Delta_{2}\). Panels (a) and (c) display experimental results, while (b) and (d) show simulations under STD. The red star and blue dot in (a) indicate the operating points for the PSE-CZ and traditional CZ gates, respectively. In (b)-(d), \textbf{PC} contour lines are superimposed for reference.}
        \label{fig:3}
    \end{figure}

    To do so, we map the leakage and CPhase landscapes in the expanded $\Delta_1$-$\Delta_2$ parameter space.
    Experimentally, we initialize the system in \(|11\rangle\) and quantify both leakage and the conditional phase via Ramsey interferometry after the gate operation.
    As shown in Fig.~\ref{fig:3}, the experimental maps and their corresponding theoretical simulations show excellent agreement, validating our physical model.
    The key to the PSE protocol lies in manipulating two distinct features of this space: the \textbf{LC} exhibits dual cyclic solutions whose connecting line is nearly perpendicular to the \textbf{PC} contour. While hardware imperfections misalign these features, the coupling strength $g$ acts as the crucial third control to tune their relative positioning, enabling a deliberate intersection of the two solution sets and thereby realizing an ideal CZ operation.
    Experimental results in Fig.~\ref{fig:3} directly visualize this outcome.
    The PSE-CZ gate (red star) successfully navigates to this optimal operating point where both conditions are met.
    In stark contrast, the traditional CZ pulse (blue dot) is restricted to the point $\Delta_1=\Delta_2=\Delta$, where a simultaneous solution for the \textbf{LC} and \textbf{PC} does not exist in the presence of hardware imperfections, a failure confirmed by simulations in Supplemental Material E. 

    By introducing an additional frequency-tuning freedom to expand the solution space, the PSE-CZ protocol effectively mitigates errors from hardware-induced non-cyclicity.

    \begin{figure}
        \centering
        \includegraphics[width=1\linewidth]{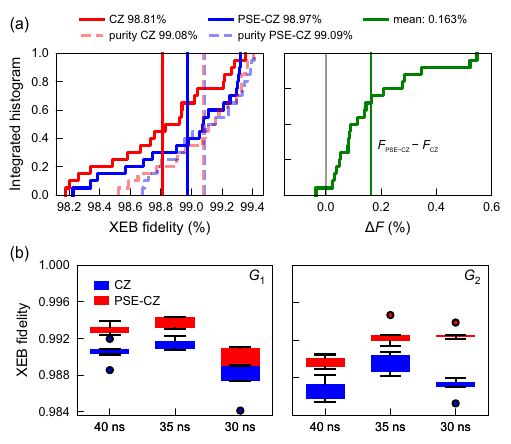}
        \caption{\textbf{Cross-entropy benchmarking (XEB) results.} (a) Cumulative distribution functions of XEB fidelity (left) 
        and fidelity improvement $\Delta F = F_{\rm PSE\text{-}CZ} 
        - F_{\rm CZ}$ (right) across 20 qubit pairs at a fixed gate duration of $50\,\mathrm{ns}$ on the 72-qubit 
        processor ``Wukong.'' 
        This duration was chosen to be compatible with the attainable coupling strengths of the measured qubit pairs. 
        The XEB purity, which characterizes the 
        decoherence-limited fidelity, is shown as lighter-colored dashed 
        lines alongside the XEB fidelity for both PSE-CZ and traditional 
        CZ gates.
        (b) XEB fidelity for qubit pairs $G_1$ and $G_2$ at gate 
        durations of $40\,\mathrm{ns}$, $35\,\mathrm{ns}$, and 
        $30\,\mathrm{ns}$. 
        In the box plots, dot symbols indicate statistical outliers 
        and whiskers represent the range of non-outlier data.}
        \label{fig:4}
    \end{figure}

    To quantify the overall performance gain, we performed cross-entropy benchmarking (XEB) on 20 qubit pairs at a fixed gate duration of $50\,\mathrm{ns}$ [Fig.~\ref{fig:4}(a)]. 
    The PSE-CZ gate achieves a mean fidelity improvement of $0.163\%$ over the traditional CZ. 
    Purity measurements (see Supplemental Material~H for methodology) confirm that both gates are dominated by decoherence errors ($\sim 0.9\%$), but the PSE protocol reduces the residual coherent error from $0.27\%$ to $0.12\%$. 
    Based on Fig.~\ref{fig:2}(c), the calculated fidelity loss from $|20\rangle$ subspace leakage (see Supplemental Material~H) drops from $0.12\%$ to $0.04\%$. 
    The remaining coherent error could attributable to leakage into coupler modes \cite{sung2021realization} and residual parasitic couplings inherent to large-scale processors \cite{zajac2021spectatorerrorstunablecoupling, ni2022scalable}, which are not specifically addressed by the PSE protocol.
    To evaluate the robustness of our protocol across different speeds, we measured gate fidelities for durations of $30$, $35$ and $40\,\mathrm{ns}$ [Fig.~\ref{fig:4}(b)]. 
    At every sampled point, the PSE-CZ gate consistently maintains a higher fidelity than the traditional CZ gate. 
    This persistent performance gap confirms that traditional symmetric pulses are unable to fully suppress distortion-induced coherent errors at these shortened interaction windows. 
    By contrast, the superior performance of the PSE scheme demonstrates its effectiveness in mitigating short-term distortions that otherwise limit gate fidelity. 
    These results confirm that the parameter-space expansion successfully circumvents the speed-fidelity trade-off, enabling fast and high-fidelity operations beyond the reach of conventional methods.

    In summary, we have identified a key physical mechanism underlying the speed–fidelity trade-off in fast cyclic quantum gates. We show that hardware-induced short-timescale distortions break the time-reflection symmetry of control pulses, rendering conventional symmetric protocols intrinsically over-constrained in the fast-gate regime, and thereby contributing to the observed trade-off.
    To overcome this limitation, we introduce the Parameter-Space-Expansion (PSE) control strategy. By judiciously incorporating an auxiliary degree of freedom, PSE restores controllability and enables cyclic closure without extending the gate duration. Our experimental implementation on a superconducting processor demonstrates that this approach can effectively suppress coherent errors from $0.27\%$ to $0.12\%$, bringing fast CZ gate performance close to the decoherence limit.    
    Looking forward, the PSE framework provides a general paradigm for robust cyclic quantum control that extends beyond a specific hardware platform. Because it addresses a symmetry-breaking mechanism inherent to cyclic evolution, this approach applies broadly to geometric (holonomic) quantum gates across platforms, including trapped-ion and solid-state systems. These results establish a practical route toward fast and high-fidelity quantum gates under realistic conditions.

	We thank Prof.~Chang-Ling Zou for reviewing the manuscript and providing valuable suggestions. This work is supported by the National Natural Science Foundation of China (Grants No.~12034018 and No.~11625419). This work is partially carried out at the USTC Center for Micro and Nanoscale Research and Fabrication.

    The authors have no conﬂicts to disclose.
        
    The data that support the ﬁndings of this study are available from the corresponding author upon reasonable request.

\bibliography{ref}

\end{document}


\title{Supplementary Information: Fast High-Fidelity CZ Gates with Robustness to Short-Term Waveform Distortion}
    
    \author{Ze-An Zhao}\addressa\addressb
    \author{Hai-Feng Zhang}
    \author{Tian-Le Wang}
    \author{Xiao-Yan Yang}
    \addressa\addressb
    \author{Peng Wang}
    \addressa\addressb\addressd
    \author{Ren-Ze Zhao}
    \addressa\addressb
    \author{Sheng Zhang}
    \addressa\addressb\addressd
    \author{Zhi-Fei Li}
    \author{Yuan Wu}
    \author{Zi-Hao Fu}
    \author{Sheng-Ri Liu}
    \addressa\addressb
	\author{Peng Duan}
    \email{pengduan@ustc.edu.cn}
    \thanks{Corresponding author}
    \addressa\addressb

    \author{Guo-Ping Guo}
    \email{gpguo@ustc.edu.cn}
    \thanks{Corresponding author}
    \addressa\addressb\addressc

\setcounter{secnumdepth}{2}
\renewcommand{\theequation}{S\arabic{equation}}
\renewcommand{\thefigure}{S\arabic{figure}}
\renewcommand{\thetable}{S\arabic{table}}
\renewcommand{\bibnumfmt}[1]{[S#1]}
\renewcommand{\citenumfont}[1]{S#1}
\setcounter{figure}{0}
\setcounter{table}{0}
\setcounter{equation}{0}

\renewcommand\thesection{S\arabic{section}}      
\renewcommand\thesubsection{\Alph{subsection}}

%

\maketitle
 \onecolumngrid
\addcontentsline{toc}{section}{Supplementary Material}

This supplement offers an elaboration of the claims made in the main text. Specifically, Section A details the system's wiring circuits. Section B presents the pertinent parameters of the chips involved. In Section C, we provide additional data on the distortion test results for both the qubits and couplers. Section D offers a theoretical exposition of the symmetry between the Hamiltonian and constraints under cyclic control. Sections E simulates the behavior of two-qubit gates in the presence of distortion. Section F shows how to amplify phase errors and leakage. Section G presents the multi-level Hamiltonian of the two transmon qubits and provides a quantitative justification for the reduction to the effective two-level $|11\rangle$--$|20\rangle$ subspace underlying the CZ gate mechanism. Section H characterizes the CZ gate errors, employing purity XEB to quantify decoherence-induced errors and analyzing the impact of 
leakage to the $|20\rangle$ state on gate fidelity.

\subsection{\label{app:subsec}System's wiring information}

\begin{figure}[htb]
    \centering
    \includegraphics[width=0.8\linewidth]{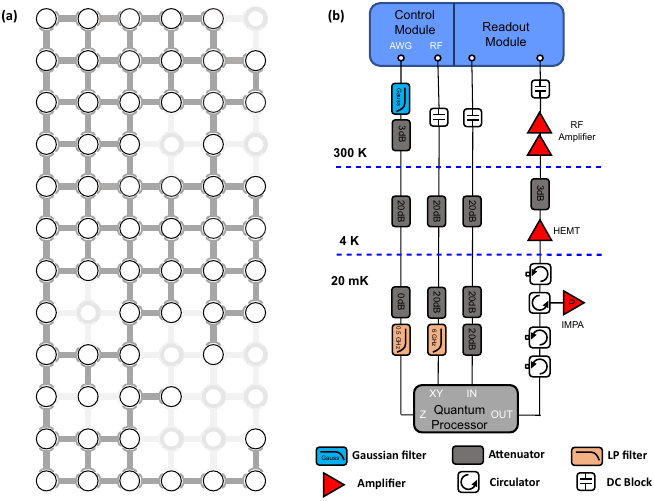}
    \caption{(a) A schematic diagram of the 72 bit superconducting quantum processor. (b) The complete wiring diagram of the experimental device.}
    \label{fig:s1}
\end{figure}

A two-dimensional (2D) flip-chip superconducting quantum processor is used in this study. Six randomly selected qubit pairs, each comprising two transmon qubits coupled by a tunable coupler, were employed to compare the traditional CZ gate with the PSE-CZ scheme. Single-qubit rotations and frequency modulation were achieved using microwave (XY) pulses and flux (Z) pulses on the individual control lines of the qubits. The coupling strength of each qubit pair was adjusted via a dedicated flux (Z) control line on the couplers. Further details of the experimental setup, including diagram of the 72 bit superconducting quantum processor "WuKong" and wiring diagram, are provided in Fig.~\ref{fig:s1}.

The Z-control signal for the qubit is generated by an arbitrary waveform generator (AWG), capable of superimposing any arbitrary waveform onto a voltage bias. The XY-control signals for the qubit are produced using a mixing module. The readout signal is amplified at the mixing chamber (MC) stage with impedance-matched parametric amplifier (IMPA)~\cite{duan2021broadband}. As illustrated in Fig.~ \ref{fig:s1}(b), attenuators are configured at different temperature stages of the cryostat to mitigate noise effects on the qubit.

\subsection{\label{app:subsec2}Overview of two-qubit gates structure}

\begin{figure*}[htb]
\centering
    \includegraphics[width=0.75\linewidth]{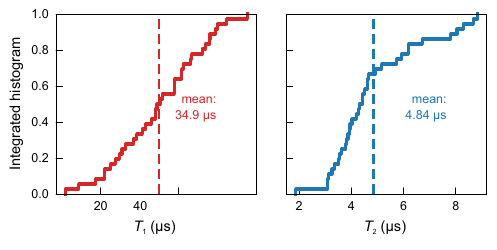}
    \caption{The cumulative distribution functions of $T_1$ and $T_2$ for the 20 qubit pairs. }
    \label{fig:s2}
\end{figure*}

Our quantum processor consists of tunable qubits (\(\omega_j\)) and tunable couplers, enabling both the effective suppression of residual coupling and the maintenance of high coupling strength, thereby facilitating fast and high-fidelity two-qubit gates.  The interaction Hamiltonian of the system can be approximated as follows,
\begin{equation}
\tilde{H}=\sum_{j=1,2} \frac{1}{2} \tilde{\omega}_{j} \sigma_{j}^{z}+\left[\frac{g_{1} g_{2}}{\Delta}+g_{12}\right]\left(\sigma_{1}^{+} \sigma_{2}^{-}+\sigma_{1}^{-} \sigma_{2}^{+}\right) ,
\nonumber
\end{equation}
where  \(\tilde{\omega}_{j}=\omega_{j}+g_{j}^{2} / \Delta_{j}\)  is the Lamb-shifted qubit frequency and  \(1 / \Delta=\left(1 / \Delta_{1}+1 / \Delta_{2}\right) / 2<0\), \(g_{j}\) represent coupling between qubit and coupler, while \(g_{12}\)  represent coupling between qubits. Effective coupling strength between qubits \(g=\left[\frac{g_{1} g_{2}}{\Delta}+g_{12}\right]\) is tunable from 0 MHz to more than 60 MHz. The coupling can be turned off during single-qubit gate operation and measurement, mitigating crosstalk from residual coupling. Compared to fixed-frequency chips, the fidelity of today's two-qubit gates is primarily limited by faster decoherence. 

As an example, the typical parameters for the two-qubit gates are detailed in Table~\ref{tab:1}, while the cumulative distribution functions of $T_1$ and $T_2$ for the 20 qubit pairs are illustrated in Fig.~\ref{fig:s2}.

\begin{table}[htb]
\centering

\begin{tblr}{colspec={Q[5cm,c] Q[1cm,c] Q[1cm,c] Q[1cm,c]},
              rowsep=1pt,
              colsep=1pt}
\hline
Parameter &  $q_\text{L}$ &  $q_\text{C}$ & $q_\text{H}$ \\ \hline
$\omega/2\pi$ sweetspot (GHz) & 4.31 & 6.29 & 4.73 \\
$\omega/2\pi$ operating-point (GHz) & 4.29 & 4.92 & 4.52 \\
$\eta/2\pi$ (MHz) & -237 & -118 & -234 \\
$T_1$ operating-point ($\mu$s) & 18.1 &  & 15.0 \\
$T_2^*$ operating-point ($\mu$s) & 2.0 &  & 2.5 \\
\end{tblr}

\begin{tblr}{colspec={Q[5cm,c] Q[3cm,c]},
              rowsep=1pt,
              colsep=1pt}
$g/2\pi$ Coupling strength (MHz) &  \hspace{0em} 0 - 60 \\
\hline
\end{tblr}
\caption{The high frequency bit, low frequency bit and coupler in the two-bit gate set are taken as the unit to show the specific parameters of our quantum processor. The frequency parameters at the sweet spot and the corresponding two-qubit operating-point are listed, along with the large anharmonicities (\(\eta\)) based on the transmon structure. The $T_1$ , $T_2^*$ data at the operating-point are also presented, as well as the range of effective coupling strength between qubit. }
\label{tab:1}
\end{table}

\subsection{\label{app:subsec3}Distortion and calibration}

The flux pulse applied to a qubit is often distorted due to the characteristics of the control electronics and the qubit's environment, which can be modeled as a linear time-invariant (LTI) system. We use qubit as a distortion detector to detect and calibrate waveform distortion at low temperatures. This distortion can be characterized by the system's impulse response \(h\left ( t \right ) \). To mitigate distortion, we apply a pre-distortion to the target pulse \(\Phi_{target} \left ( t \right ) \) using an inverse impulse response \(h^{-1}\left ( t \right )  \) designed to counteract \(h\left ( t \right ) \).

\begin{figure*}
    \centering
    \includegraphics[width=0.75\linewidth]{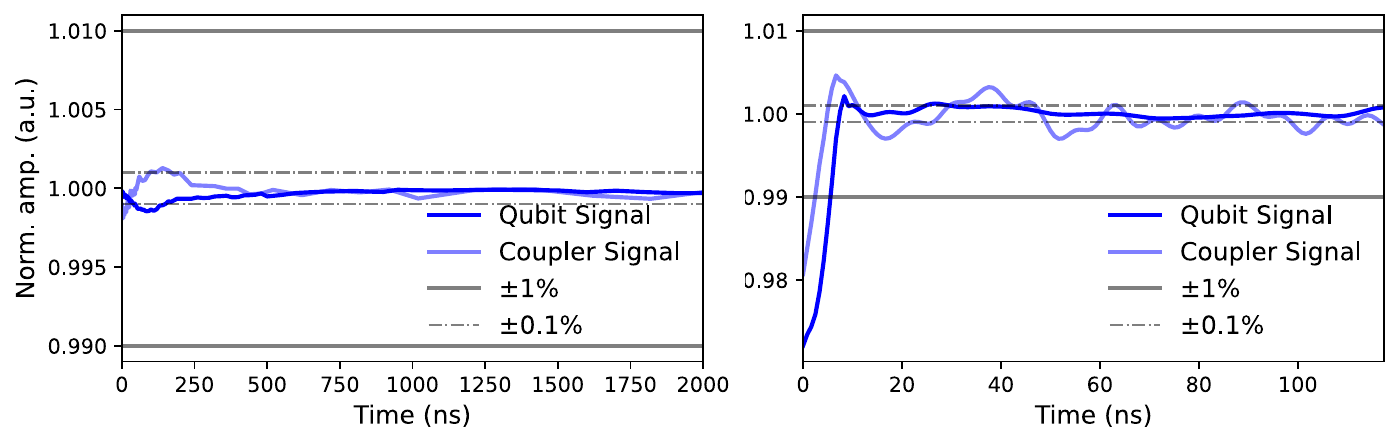}
    \caption{Step responses of the qubit and coupler, following the application of distortion corrections, were measured sequentially using the Martinis technique~\cite{foxen2018high} and DiCarlo technique~\cite{rol2020time}. We find that the coupler exhibits larger deviations and variances in the residual distortion results, indicating more severe residual distortion and a correspondingly reduced confidence in the characterization of the true response.
}
    \label{fig:s3}
\end{figure*}

The actual pulse \(\Phi\left ( t \right ) \) experienced by the qubit is given by the convolution of the applied voltage waveform from the arbitrary waveform generator (AWG) \(V_{AWG} \left ( t \right ) \) with the system's impulse response \(h\left ( t \right ) \):

$$\Phi\left ( t \right ) =\left ( h* V_{AWG}\right ) \left ( t \right ) =\left ( h*\left ( h^{-1}* \Phi_{target}\right )  \right )\left ( t \right )$$
$$=\left ( \left ( h^{-1}* h \right ) *\Phi_{target} \right ) \left ( t \right ) ,$$
where \(*\) denotes convolution.

After applying the pre-distortion, any remaining distortion is quantified by measuring the step response \(s\left ( t \right ) \) of the system, which is defined as:

\[s\left ( t \right ) =\int_{0}^{t} h^{-1}*h\left ( t_{0}  \right ) dt_{0} . \]
In the case of perfect distortion corrections, the normalized amplitude would have value 1 for all times larger than zero.

The distortion response measurement results for qubit and coupler are presented in Fig.~\ref{fig:s3}. The left and right panels show the distortion responses measured using two different methods over long and short intervals, respectively. The experimental results reveal that achieving perfect distortion calibration is challenging, particularly in time ranges shorter than \( 10\,\mathrm{ns} \). For the coupler, the discrepancy between the experimental results and the ideal square wave is larger, indicating more pronounced residual distortion compared to the qubit.

\subsection{\label{app:subsec4}Symmetry and Constraints in Cyclic Evolution}

Here, we provide the formal proofs for the relationship between time-reflection symmetry (TFS) and the number of cyclic constraints, and for the origin of the dual solutions mentioned in the main text.

\subsubsection{TFS and Propagator Symmetry}
We first prove that for a time-dependent, real, symmetric Hamiltonian $H(t)=H^T(t)$ that preserves TFS, i.e., $H(t)=H(\tau-t)$, the total propagator $U(\tau,0)$ is symmetric.

The propagator is the time-ordered product of infinitesimal evolution operators: $U(\tau, 0) = \lim_{N\to\infty} U_N U_{N-1} \cdots U_1$, where $U_n = \exp[-i\tau/\hbar N \cdot H(n\tau/N)]$.
Due to the Hamiltonian symmetry, each infinitesimal step is symmetric: $U_n = U_n^T$.
Furthermore, TFS implies $H(n\tau/N) = H((\tau - n\tau/N)) = H((N-n)\tau/N)$, which means the infinitesimal steps are symmetric in time around the midpoint: $U_n = U_{N-n+1}$.

The transpose of the total propagator is $U^T(\tau,0) = U_1^T \cdots U_N^T = U_1 \cdots U_N$. Using the time symmetry of the steps, this becomes:
\begin{align*}
    U^T(\tau,0) &= (U_{N}) (U_{N-1}) \cdots (U_1) \\
               &= U_1 U_2 \cdots U_N \\
               &= U_N^T U_{N-1}^T \cdots U_1^T \\
               &= U_N U_{N-1} \cdots U_1 = U(\tau,0). \label{eq:U_is_symmetric_TFS} \tag{S1}
\end{align*}
This confirms that TFS in a real, symmetric Hamiltonian imposes the symmetry $U(\tau,0) = U^T(\tau,0)$ on the propagator.

\subsubsection{Constraint Counting in Two and D Dimensions}
For a two-level system (qubit), the $\mathrm{SU}(2)$ propagator is $U = \begin{pmatrix} a & b \\ -b^* & a^* \end{pmatrix}$, with the normalization condition $|a|^2+|b|^2=1$. A cyclic evolution requires the off-diagonal element to vanish ($b=0$), which imposes two real constraints: $\mathrm{Re}[b]=0$ and $\mathrm{Im}[b]=0$.
If the system possesses TFS, Eq.~\eqref{eq:U_is_symmetric_TFS} imposes $U=U^{\mathrm{T}}$, which implies $b=-b^*$ or $\mathrm{Re}[b]=0$. Thus, only one constraint, $\mathrm{Im}[b]=0$, remains.

This principle generalizes to an $D$-level system. A general propagator $U \in \mathrm{SU}(D)$ can be parameterized via its generators $T_k$:
\[
U = \exp\left(-i\sum_{k=1}^{D^2-1} \theta_k T_k\right).
\]
The $\det[U]=1$ condition is inherent to the SU(D) generators, leaving the $D^2-1$ real coefficients $\theta_k$ as the independent parameters, or degrees of freedom (DOFs). The conditions for a cyclic evolution are:
\begin{enumerate}
    \item Off-diagonal nullification: $U_{ij}=0$ for $i \neq j$.
    \item Diagonal unitarity: $|U_{ii}|=1$ for all $i$.
\end{enumerate}
A matrix satisfying these conditions is diagonal. After factoring out a physically irrelevant global phase, the evolution is described by only $D-1$ independent relative phases. Therefore, a cyclic evolution reduces the system's DOFs to $D-1$.

We can now count the constraints by calculating the reduction in DOFs:
\begin{itemize}
    \item \textbf{General System (No TFS):} The propagator is a general SU(N) matrix with $D^2-1$ DOFs. The number of constraints is $(D^2-1) - (D-1) = \mathbf{D(D-1)}$.

    \item \textbf{TFS System:} The propagator is a symmetric SU(D) matrix ($U=U^{\mathrm{T}}$). The symmetry condition reduces the number of independent parameters to $\frac{D(D+1)}{2}-1$. The number of constraints is therefore $(\frac{D(D+1)}{2}-1) - (D-1) = \mathbf{\frac{D(D-1)}{2}}$.
\end{itemize}
This confirms that for any dimension $D$, breaking TFS doubles the number of constraints required to achieve a cyclic evolution.

\subsubsection{Origin of Dual Cyclic Solutions}
For a pulse that is symmetric around its midpoint (a property enforced by TFS), the evolution can be split into two halves: $U(\tau,0) = U(\tau, \tau/2) U(\tau/2,0)$. The symmetry implies $U(\tau/2,0) = U^T(\tau, \tau/2)$.
Let us parameterize the first half as $U(\tau/2,0) = \begin{pmatrix} c & d \\ -d^* & c^* \end{pmatrix}$. Then the second half is $U(\tau, \tau/2) = \begin{pmatrix} c & -d^* \\ d & c^* \end{pmatrix}$.
The total evolution is:
\[
U(\tau,0) = \begin{pmatrix} c & -d^* \\ d & c^* \end{pmatrix} \begin{pmatrix} c & d \\ -d^* & c^* \end{pmatrix} = \begin{pmatrix} c^2+d^{*2} & cd-c^*d^* \\ c^*d-cd^* & c^{*2}+d^2 \end{pmatrix}.
\]
The cyclic condition requires the off-diagonal element to be zero: $cd-c^*d^*=0$. This is equivalent to $\mathrm{Im}[cd^*]=0$. If we write $c = |c|e^{i\phi_c}$ and $d = |d|e^{i\phi_d}$, this condition becomes $\sin(\phi_c - \phi_d) = 0$, which yields:
\begin{equation}
\phi_c - \phi_d = M\pi, \quad \text{where } M=0 \text{ or } 1. \label{eq:dual_solutions} \tag{S2}
\end{equation}
These two distinct families of solutions, corresponding to $M=0$ and $M=1$, are the dual cyclic solutions referred to in the main text.

\subsection{\label{app:subsec5}
Theoretical Fidelity of Two-Qubit Gates Simulated with Distortion
}

The actual CZ gate control waveform simplifies to: Applying frequency detuning exclusively to the high-frequency qubit while maintaining the low-frequency qubit fixed, with separate modulation of the inter-qubit coupling strength \(g(t)\). Both parameters are modeled as flat-top Gaussian waveforms in this study. By applying butterworth low-pass filters to these waveforms, we simulate the effects of various distortions by adjusting the filter parameters in gate duration of \( 40\,\mathrm{ns} \). The theoretical fidelity $F$ is then computed within the computational subspace as a function of $\Delta(t)$ and $g(t)$.

\begin{figure*}
    \centering
    \includegraphics[width=0.66\linewidth]{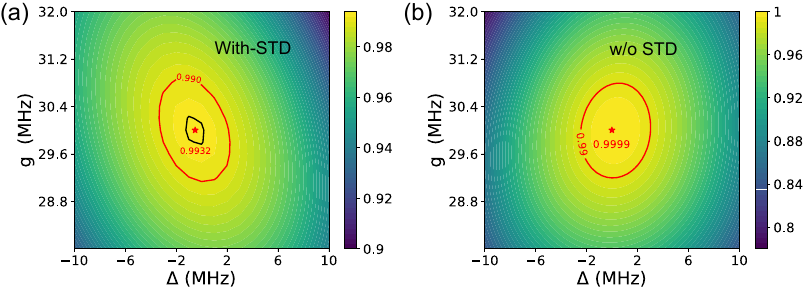}
    \caption{(a),(b) Simulated result of fidelity with and without (w/o) STD as \(\Delta\) and \(g\) are varied, with the maximum fidelity marked by a red star.}
    \label{fig:s4}
\end{figure*}

\begin{figure}
    \centering
    \includegraphics[width=0.66\linewidth]{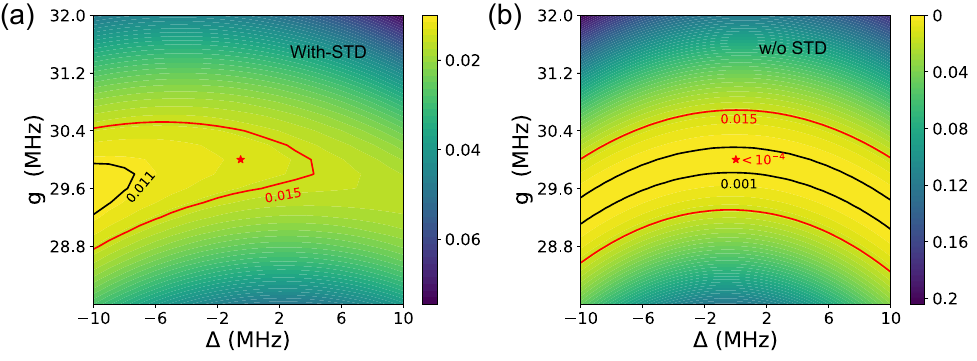}
    \caption{(a),(b) Simulated leakage with and without STD as functions of $\Delta$ and $g$. The red star indicates the maximum fidelity and corresponds to the same point as in Fig.~\ref{fig:s4}.
}
    \label{fig:s5}
\end{figure}

The CZ gate fidelity can be computed using the following formula:

\[F = \frac{1}{n(n+1)} \left[ \text{Tr}(MM^\dagger) + \left| \text{Tr}(M) \right|^2 \right],\]

where $ M = U_0^\dagger U $, $ U $ is the actual evolution matrix, and $ U_0 $ is the ideal evolution matrix. Since $ M $ is unitary, $\text{Tr}(MM^\dagger) = n$.

To simulate the effects of short-term distortion (STD), we model the system's response using a digital low-pass filter. This filter is applied 
as a numerical post-processing step in simulation and does not represent 
a hardware component in the experimental setup. Specifically, we employ the Butterworth filter implementation from the \texttt{scipy.signal} library, which is characterized by three main parameters: the filter order \(n\), the sampling rate \(f_s\), and the cutoff frequency \(f_c\). The filter is defined using a normalized cutoff frequency, calculated as \(f_c / (0.5 \cdot f_s)\). For the simulations presented in this letter, we used a configuration with a 4th-order filter (\(n=4\)), a sampling rate of \(f_s = 50~\mathrm{GHz}\), and a cutoff frequency of \(f_c \approx 280~\mathrm{MHz}\).

To control the magnitude of the distortion, we introduce a dimensionless scaling parameter, \(\alpha\). The final distorted waveform, \(W_{\text{distorted}}(t)\), is constructed by taking the difference between the ideal waveform, \(W_{\text{ideal}}(t)\), and the filtered waveform, \(W_{\text{filtered}}(t)\), scaling this difference by \(\alpha\), and adding it back to the ideal waveform:
\[
W_{\text{distorted}}(t) = W_{\text{ideal}}(t) + \alpha \cdot \left( W_{\text{filtered}}(t) - W_{\text{ideal}}(t) \right).
\]
This model allows us to smoothly vary the distortion strength. A value of \(\alpha=0\) corresponds to an ideal, undistorted pulse, while \(\alpha=1\) represents the full effect of the low-pass filter.

This filtering process produces a distortion response with a characteristic timescale of approximately $10~\mathrm{ns}$, which is consistent with the timescales that are challenging to fully characterize and correct in experimental distortion calibration. As illustrated in Fig.~ \ref{fig:s4}(a,b), this model predicts a minimal coherent error of 0.68\% for the traditional gate under distortion. Furthermore, the simulated gate error escalates with increasing \(\alpha\), corresponding to amplified distortion amplitudes, which is in qualitative agreement with the experimental trends shown in Fig.1(c) of the main text.

The optimal working point with the highest fidelity, indicated by the red star in Fig.~\ref{fig:s5}, does not coincide with the region of minimal leakage. This suggests that under distortion, the traditional CZ pulse cannot simultaneously satisfy \textbf{LC} and \textbf{PC}, consistent with the theoretical analysis presented in the text.

\subsection{\label{app:subsec6}Error amplification}

Any evolution in the \( |11\rangle -|20\rangle \) subspace can be represented by an element of $\mathrm{SU}(2)$:
$$
R(\Omega, \zeta, \chi) = \begin{pmatrix}
e^{-i\zeta}\cos\Omega & -ie^{i\chi}\sin\Omega \\
-ie^{-i\chi}\sin\Omega & e^{i\zeta}\cos\Omega
\end{pmatrix}.
$$
Setting $\chi = 0$ to eliminate measurement-induced phase effects, we obtain:
$$
R(\Omega, \zeta) = \begin{pmatrix}
e^{-i\zeta}\cos\Omega & -i\sin\Omega \\
-i\sin\Omega & e^{i\zeta}\cos\Omega
\end{pmatrix} = e^{-i\theta \hat{\mathbf{n}} \cdot \boldsymbol{\sigma}},
$$
where $\boldsymbol{\sigma} = (\sigma_x, \sigma_y, \sigma_z)$ and $\hat{\mathbf{n}} = (n_x, n_y, n_z)$ is a unit vector ($\|\hat{\mathbf{n}}\| = 1$). 

In the Bloch sphere representation, $\hat{\mathbf{n}}$ defines the rotation axis, and $\theta = \arccos(\cos\Omega\cos\zeta)$ quantifies the rotation angle. The rotation matrix is expressed as: 
$$
R(\theta, \alpha) = e^{-i\theta \hat{\mathbf{n}} \cdot \boldsymbol{\sigma}},\quad
\text{with} \quad \hat{\mathbf{n}} \cdot \boldsymbol{\sigma} = \begin{pmatrix} \cos\alpha & \sin\alpha \\ \sin\alpha & -\cos\alpha \end{pmatrix}.
$$
For simplicity and without loss of generality, we align the coordinate system such that $n_y = 0$. This yields $\alpha = \arctan(\tan\Omega/\sin\zeta)$. 

Experimentally, parameters satisfying the leakage condition \textbf{LC}  ($\theta =\pi + \delta, \delta \approx 0$) are first identified to minimize leakage. Next, parameters fulfilling the CPhase condition \textbf{PC} for the  \( |11\rangle \) state are optimized. Theoretically, both \( |11\rangle \) and \( |20\rangle \) acquire a $-1$ phase , which enforces $\alpha \approx 0$ in $R(\theta, \alpha)$(ignoring a -1 global phase).

\subsubsection*{1. CPhase error amplification:}
CPhase error can be amplified via repeated CZ gate applications.
The probability of measuring the $|11\rangle$ state after repeating 
the CZ gate $n_{\rm CZ}$ times is given by:
\begin{equation*}
P_{11}(n_{\rm CZ}) = |\langle 11 | R(\theta, \alpha)^{n_{\rm CZ}} 
| 11 \rangle|^2 = |\cos(n_{\rm CZ}\theta) - i \sin(n_{\rm CZ}\theta) 
\cos\alpha|^2.
\end{equation*}
It is straightforward to show that $P_{11}(n_{\rm CZ})\approx 1$ for 
$\alpha \approx 0$, indicating that leakage does not accumulate with 
$n_{\rm CZ}$. However, the phase error, given by:
\begin{equation*}
\varphi_{\rm err}=\arg\left(\langle 11 | R(\theta, \alpha)^{n_{\rm CZ}} 
| 11 \rangle\right) - n_{\rm CZ}\pi \approx -n_{\rm CZ}\theta - 
n_{\rm CZ}\pi = -n_{\rm CZ}\delta,
\end{equation*}
accumulates linearly with $n_{\rm CZ}$, as shown in Fig.2(c) bottom.

\subsubsection*{2. Leakage amplification:}
By strategically modulating intergate delays at idle working points, 
we enhance leakage into the $|20\rangle$ state, which is subsequently 
detected using a two-state readout technique. This delay modulation at 
the idle working points controls the relative phase between $|11\rangle$ 
and $|20\rangle$ states, thereby coherently enhancing the leakage error 
through constructive interference.

Noting that $R(\Omega, \zeta)=Z(\zeta)R_X(\Omega)Z(\zeta)$, inserting 
phase gate $Z(-2\zeta)$ between CZ gates yields:
\begin{equation*}
\left[R(\Omega, \zeta)Z(-2\zeta)\right]^{n_{\rm CZ}-1}R(\Omega, \zeta)
=Z(\zeta)R_X(n_{\rm CZ}\,\Omega)Z(\zeta).
\end{equation*}
By tuning the relative phase between $|11\rangle$ and $|20\rangle$ via 
$Z(-2\zeta)$, the composite operation simplifies to an effective 
$X$-rotation. This maximizes constructive interference, enabling rapid 
and coherent amplification of the leakage population.

The experimental signal is fitted using:
\begin{equation*}
f(n_{\rm CZ}) = C\!\left(1 - e^{-n_{\rm CZ}/\tau_{\rm relax}}\right) 
+ A\,e^{-n_{\rm CZ}/\tau_{\rm osc}}\cos(\omega\, n_{\rm CZ}+\phi) 
- A\cos(\phi),
\end{equation*}
where $n_{\rm CZ}$ denotes the number of gate repetitions, and $C$ and 
$A$ are constants. The parameters $\tau_{\rm relax}$ and $\tau_{\rm osc}$ 
represent the relaxation and dephasing time constants, respectively, 
while $\phi$ is the initial phase. The single-gate leakage is extracted 
from the fitted oscillation frequency $\omega$ via $L = \sin^2(\omega/2)$, 
following directly from the SU(2) rotation structure in the 
$|11\rangle$--$|20\rangle$ subspace. This frequency-based extraction 
is robust against noise and slow drift.

\subsection{\label{app:subsec7}Multi-level Structure and Reduction to the Effective Two-level Subspace}

The physical mechanism of the CZ gate involves the multi-level
structure of the two transmon qubits. Within the rotating-wave
approximation, excitation-number-conserving dynamics decouple the
system into independent manifolds. The full Hamiltonian in the
basis $\{|00\rangle, |01\rangle, |10\rangle, |11\rangle,
|20\rangle, |02\rangle\}$ takes the block-diagonal form:

\begin{equation*}
H = \begin{pmatrix}
0 & 0 & 0 & 0 & 0 & 0 \\
0 & \omega_2 & g_1 & 0 & 0 & 0 \\
0 & g_1 & \omega_1 & 0 & 0 & 0 \\
0 & 0 & 0 & \omega_1+\omega_2 & g_2 & g_2 \\
0 & 0 & 0 & g_2 & 2\omega_1+\eta_1 & 0 \\
0 & 0 & 0 & g_2 & 0 & 2\omega_2+\eta_2
\end{pmatrix},
\end{equation*}
where $g_1$ denotes the effective coupling between $|01\rangle$
and $|10\rangle$ in the single-excitation manifold, and $g_2$
denotes the effective coupling between $|11\rangle$ and $|20\rangle$
(as well as between $|11\rangle$ and $|02\rangle$) in the
double-excitation manifold. The vacuum state $|00\rangle$ is
decoupled from all other states and serves as the energy reference.
We consider the relevant states in the single-excitation manifold
($|01\rangle$, $|10\rangle$) and the double-excitation manifold
($|11\rangle$, $|20\rangle$, $|02\rangle$), and show that the
dynamics reduce to the two-level $|11\rangle$--$|20\rangle$
subspace analyzed in the main text.

\textbf{Single-excitation manifold.}
The states $|01\rangle$ and $|10\rangle$ are coupled through the
tunable coupler with an effective coupling strength $g_1$.
The CZ gate is activated by bringing $|11\rangle$ and $|20\rangle$
into near-resonance, which requires $\omega_1 + \eta_1 \approx
\omega_2$, i.e., the two qubits are detuned by approximately
$|\omega_1 - \omega_2| \approx |\eta_1| \approx 230$--$240~\mathrm{MHz}$.
In our standard CZ gate implementation, the effective coupling
strength between $|11\rangle$ and $|20\rangle$ is
$g_2 \approx 16.7~\mathrm{MHz}$; at the same coupler bias point,
the effective coupling $g_1$ in the single-excitation manifold
is smaller than $g_2$ (see Ref.~\cite{sung2021realization}, Fig.~2), i.e.,
$g_1 < 16.7~\mathrm{MHz}$.
Since the detuning between $|01\rangle$ and $|10\rangle$ satisfies
$|\omega_1 - \omega_2| \approx 230~\mathrm{MHz} \gg g_1$,
no appreciable population transfer occurs in the single-excitation
manifold during the CZ gate.

\textbf{Double-excitation manifold.}
In the double-excitation manifold, the state $|02\rangle$ is detuned
from $|11\rangle$ by:
\begin{equation*}
    \Delta_{02} \equiv (2\omega_2 + \eta_2) - (\omega_1 + \omega_2)
    = (\omega_2 - \omega_1) + \eta_2.
\end{equation*}
Using the near-resonance condition $\omega_2 - \omega_1 \approx
-\eta_1$ and the fact that both qubits have similar anharmonicities
$\eta_1 \approx \eta_2 \approx \eta$, we obtain:
\begin{equation*}
    |\Delta_{02}| \approx |\eta_1 + \eta_2|
    \approx 2|\eta| \approx 460\text{--}480~\mathrm{MHz}.
\end{equation*}
This detuning is more than an order of magnitude larger than
$g_2 \approx 16.7~\mathrm{MHz}$, satisfying $|\Delta_{02}| \gg g_2$.
The population admixture of $|02\rangle$ into the gate dynamics
is therefore suppressed to order $(g_2/|\Delta_{02}|)^2 \lesssim
10^{-3}$ and is negligible.

\textbf{Reduction to effective two-level system.}
With both the $|01\rangle$/$|10\rangle$ exchange and the $|02\rangle$
coupling suppressed by large detunings, the relevant dynamics during
the CZ gate are well described by the two-level subspace
$\{|11\rangle, |20\rangle\}$. The effective Hamiltonian in this
subspace takes the form:
\begin{equation*}
    H_\mathrm{eff}(t) = \frac{1}{2}\hbar g_2(t)\,\sigma_x
    - \frac{1}{2}\hbar\Delta(t)\,\sigma_z,
\end{equation*}
where $g_2(t)$ is the time-dependent coupling strength between
$|11\rangle$ and $|20\rangle$ (denoted $g(t)$ in the main text), $\Delta(t)$ is their frequency detuning, and $\sigma_x$, $\sigma_z$ are Pauli operators in this subspace. This is the Hamiltonian adopted in the main text and forms the basis of the cyclic evolution analysis underlying the PSE-CZ protocol.

\subsection{\label{app:subsec8}Analysis of CZ gate error}

In this appendix, we describe two complementary methods used to characterize and quantify the errors of the CZ gate implemented in this work: purity cross-entropy benchmarking (purity XEB) for decoherence errors, and a leakage-based fidelity calculation for errors arising from population transfer to the $|20\rangle$ state.

\subsubsection*{1, Purity XEB for decoherence errors.}

In XEB~\cite{arute2019quantum_supremacy}, random quantum circuits are applied and the resulting bitstrings are sampled experimentally. The key observation is 
that the output of a random quantum circuit exhibits a speckle-like 
probability distribution, which is progressively washed out by errors. 
Rather than reconstructing the full output distribution, XEB uses 
numerical simulations to evaluate the ideal probabilities of the 
experimentally sampled bitstrings, from which the depolarization 
fidelity per cycle can be extracted.

While $F_{\mathrm{XEB}}$ captures the overall gate performance, it does not directly distinguish between coherent errors (e.g., control imperfections, residual $ZZ$ coupling) and incoherent errors (e.g., dephasing and relaxation). To isolate the contribution of decoherence, we employ purity XEB.

In purity XEB, after executing the standard XEB random circuit sequence, two-qubit state tomography is performed on the final state to reconstruct the density matrix $\rho$. The purity $\gamma = \mathrm{Tr}(\rho^2)$ of the output state is then extracted. For a purely coherent (unitary) error channel, the output state remains pure ($\gamma = 1$), whereas decoherence reduces the purity below unity. By comparing the purity to the XEB fidelity, one can decompose the total gate error into a coherent part and an incoherent (decoherence) part. Specifically, the decoherence error per cycle $e_{\mathrm{dec}}$ is related to the purity decay rate, while coherent errors manifest as a discrepancy between $F_{\mathrm{XEB}}$ and the purity-inferred fidelity.

This protocol allows us to attribute the dominant source of infidelity in our CZ gate to dephasing-induced decoherence, which is consistent with the $T_2$ limitations observed in our device characterization.

\subsubsection*{2, Leakage to $|20\rangle$ and gate fidelity.}

To estimate the impact of leakage to $|20\rangle$ on gate fidelity, we truncate the full evolution operator $U$ to the computational subspace $\{|00\rangle,|01\rangle,|10\rangle,|11\rangle\}$, yielding the truncated matrix $U_{\mathrm{trunc}}$. The gate fidelity relative to the ideal CZ gate $U_0$ is then evaluated as
\begin{equation*}
    F = \frac{1}{n(n+1)} \left[ \mathrm{Tr}(MM^\dagger) + \left|\mathrm{Tr}(M)\right|^2 \right],
    \label{eq:gate_fidelity}
\end{equation*}
where $M = U_0^\dagger U_{\mathrm{trunc}}$ and $n = 4$. In the absence of leakage, $U_{\mathrm{trunc}}$ is unitary and $\mathrm{Tr}(MM^\dagger) = n$; any population loss to $|20\rangle$ reduces this term below $n$, directly contributing to the infidelity $1-F$.

\bibliography{ref}